\documentclass[aps,prb,twocolumn,amsmath,amssymb,superscriptaddress]{revtex4-2}
\usepackage{graphicx}
\usepackage{amssymb}
\usepackage{color}
\usepackage{amsmath}
\usepackage{float}
\usepackage{gensymb}
\usepackage{chemformula}
\usepackage{epstopdf}
\usepackage{hyperref}
\usepackage{makecell}
\usepackage{multirow}
\usepackage{afterpage}
\usepackage{tabularx}
\usepackage{amsmath}
\usepackage{float}
\usepackage{booktabs}

\newcommand{\beq}{\begin{eqnarray}}
	\newcommand{\eeq}{\end{eqnarray}}

\hypersetup{hidelinks,
	colorlinks=true,
	allcolors=blue,
	pdfstartview=Fit,
	breaklinks=true}

\begin{document}
	\title{Signatures of three-state Potts nematicity in spin excitations of the van der Waals antiferromagnet FePSe$_3$}
	\author{Weiliang~Yao}
	\email{wy28@rice.edu}
	\affiliation{Department of Physics and Astronomy, Rice University, Houston, Texas 77005, USA}
	\affiliation{Rice Laboratory for Emergent Magnetic Materials and Smalley-Curl Institute, Rice University, Houston, Texas 77005, USA.}
	\author{Viviane~Peçanha-Antonio}
	\affiliation{ISIS Facility, STFC, Rutherford Appleton Laboratory, Chilton, Didcot, Oxfordshire OX11 0QX, United Kingdom}
	\author{Devashibhai~Adroja}
	\affiliation{ISIS Facility, STFC, Rutherford Appleton Laboratory, Chilton, Didcot, Oxfordshire OX11 0QX, United Kingdom}
	\affiliation{Highly Correlated Matter Research Group, Physics Department, University of Johannesburg, Auckland Park 2006, South Africa}
	\author{S.~J.~Gomez~Alvarado}
	\affiliation{Department of Physics and Astronomy, Rice University, Houston, Texas 77005, USA}
	\affiliation{Rice Laboratory for Emergent Magnetic Materials and Smalley-Curl Institute, Rice University, Houston, Texas 77005, USA.}
	\author{Bin~Gao}
	\affiliation{Department of Physics and Astronomy, Rice University, Houston, Texas 77005, USA}
	\affiliation{Rice Laboratory for Emergent Magnetic Materials and Smalley-Curl Institute, Rice University, Houston, Texas 77005, USA.}
	\author{Sijie~Xu}
	\affiliation{Department of Physics and Astronomy, Rice University, Houston, Texas 77005, USA}
	\affiliation{Rice Laboratory for Emergent Magnetic Materials and Smalley-Curl Institute, Rice University, Houston, Texas 77005, USA.}
	\author{Ruixian~Liu}
	\affiliation{Center for Advanced Quantum Studies and Department of Physics, Beijing Normal University, Beijing 100875, People’s Republic of China}
	\author{Xingye~Lu}
	\email{luxy@bnu.edu.cn}
	\affiliation{Center for Advanced Quantum Studies and Department of Physics, Beijing Normal University, Beijing 100875, People’s Republic of China}
	\author{Pengcheng~Dai}
	\email{pdai@rice.edu}
	\affiliation{Department of Physics and Astronomy, Rice University, Houston, Texas 77005, USA}
	\affiliation{Rice Laboratory for Emergent Magnetic Materials and Smalley-Curl Institute, Rice University, Houston, Texas 77005, USA.}
	
	\date{\today}
	
	\begin{abstract}
		In two-dimensional (2D) nearly square-lattice quantum materials, electron correlations can induce an electronic nematic phase with twofold rotational ($C_2$) symmetry that profoundly impacts their properties. For 2D materials with threefold rotational ($C_3$) symmetry, such as the honeycomb lattice, a vestigial three-state Potts nematic order has been observed in the van der Waals antiferromagnet (AFM) FePSe$_3$ via optical and thermodynamic methods under uniaxial strain. Here, we use neutron scattering to study the magnetic order and spin excitations of FePSe$_3$ under uniaxial strain. In the AFM ordered state, we find that $\sim$0.6\% tensile strain significantly suppresses one zigzag domain and promotes the other two, lowering the AFM order and spin waves to $C_2$ symmetry. The broken $C_3$ symmetry in spin excitations persists slightly above $T_{\rm{N}}\approx 108.6$ K, where the zigzag AFM order is absent. Our results thus provide direct evidence of magnetoelastic coupling and suggest that the three-state Potts nematicity in paramagnetic spin excitations arises from the vestigial order associated with the low-temperature zigzag AFM order.
	\end{abstract}
	
	\maketitle
	
	Symmetry breaking, such as the formation of new magnetic, structural, or electronic orders, is a fundamental phenomenon in nature. Understanding the nature of symmetry breaking can provide crucial insights into the underlying laws that govern the physical properties of materials \cite{BeekmanSPPLN2019}. In periodic crystalline solids, rotational symmetry refers to the property that the material remains invariant under rotation by a specific angle. Owing to translational symmetry, crystalline solids can only host twofold, threefold, fourfold, or sixfold discrete rotational symmetries, denoted as $C_2$, $C_3$, $C_4$, and $C_6$, respectively \cite{Ashcroft1976}.
	
	A nematic phase originates from a state where elongated molecules in liquid crystals exhibit no crystalline positional order but are aligned with their long axes approximately parallel, forming directional order with $C_2$ symmetry \cite{DeGennes}. In liquid crystals, it is an intermediate phase between an ordered crystalline solid and a disordered liquid \cite{DeGennes}. An electronic nematic phase, predicted to occur near a two-dimensional (2D) nearly square-lattice Mott insulator \cite{KivelsonNature1998}, typically breaks the $C_4$ symmetry of the underlying lattice and arises near a superconducting phase \cite{FradkinAnnuRevCMP2010,FernandesAnnuRevCMP2010,NiePNAS2014,DaiRMP2015,BohmerNP2022}. Since its discovery in copper- and iron-based high-temperature superconductors with nearly square-lattice structures \cite{HinkovScience2008,ChuScience2010}, studies of nematicity have often focused on the spontaneous reduction of $C_4$ symmetry to $C_2$ in transport, electronic, and magnetic properties, implying that the system selects one of two energetically equivalent configurations (Ising nematicity) \cite{FradkinAnnuRevCMP2010,FernandesAnnuRevCMP2010,NiePNAS2014,DaiRMP2015,BohmerNP2022}.
	
	For crystalline solids with $C_3$ or $C_6$ symmetry \cite{note}, it is also possible to develop a nematic state that selects one (or two) out of three equivalent configurations—a phenomenon known as three-state Potts nematicity \cite{FernandesSA2020,ChakrabortyPRB2023,XiePRB2024}. Materials in the hexagonal crystal family are thus an ideal platform to explore this type of symmetry breaking. Recent studies on systems with triangular, kagome, and honeycomb lattices have reported intriguing nematic behaviors in their electronic and/or magnetic properties, including Fe$_{1/3}$NbS$_2$ \cite{LittleNMlittle2020}, Co$_{1/3}$TaS$_2$ \cite{FengArxiv2025,KirsteinArxiv2025}, CsV$_3$Sb$_5$ \cite{NieNature2022,XuNP2022,AsabaNP2024}, ScV$_6$Sn$_6$ \cite{FarhangNC2025}, FePSe$_3$ \cite{HwangboNP2024,NiArxiv2023}, and NiPS$_3$ \cite{TanNanoLett2024}. However, studies to date have used optical, transport, or thermodynamic methods, which cannot directly probe the microscopic origin of the $C_3$ symmetry breaking in the nematic phase or determine the energy scale of the corresponding nematic fluctuations.
	
\begin{figure}[t!]
		\hspace{0cm}\includegraphics[width=8.5cm]{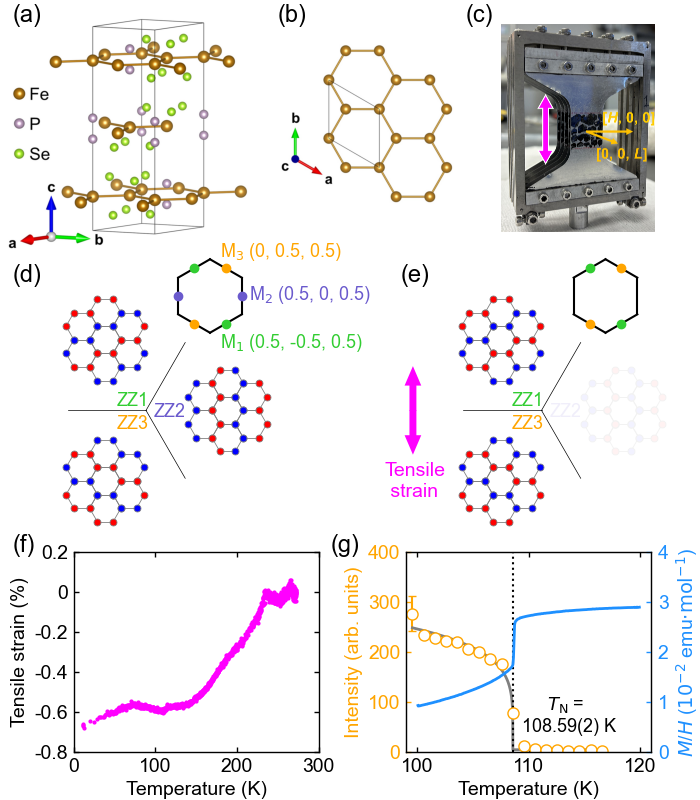}
		\caption{(a) Crystal structure of FePSe$_3$. (b) Honeycomb lattice formed by Fe ions. (c) Strain device with FePSe$_3$ single crystals attached. The magenta double arrow indicates the tensile strain direction. (d) Three zigzag domains (ZZ1, ZZ2, and ZZ3) in the unstrained Fe honeycomb lattice, where red and blue represent spins pointing into and out of the honeycomb plane. The inset (upper right) shows the six magnetic Bragg peaks in the hexagonal Brillouin zone corresponding to the three zigzag domains. (e) Two remaining zigzag domains (ZZ1 and ZZ3) under tensile strain. (f) Temperature dependence of the tensile strain applied to FePSe$_3$ single crystals \cite{MoCPL2024}. (g) Temperature dependence of the diffraction intensity at Bragg peak (0, 0.5, 0.5) (left axis) and the magnetic susceptibility under a 0.5 T $c$-axis field (right axis). The gray solid curve is a power-law fit to the diffraction data, as described in the text. The dashed vertical line marks $T_{\rm{N}}$.}
		\label{fig1}
	\end{figure}

	In this work, we focus on the van der Waals antiferromagnet FePSe$_3$, which crystallizes in a rhombohedral $R\bar{3}$ structure consisting of hexagonal layers stacked along the $c$ axis through weak van der Waals interactions [Fig. \ref{fig1}(a)] \cite{TaylorInorgChem1974,WangACSNano2022}. The Fe ions, with localized magnetic moments, form a honeycomb lattice [Fig. \ref{fig1}(b)] that develops long-range antiferromagnetic (AFM) order below $T_{\rm{N}}\approx 108$ K \cite{CuiNC2023,HwangboNP2024,NiArxiv2023,Haglund2019}. Neutron diffraction experiments identified a zigzag magnetic structure with propagation wave vector $\mathbf{k}_{\rm{m}}$ = (0.5, 0, 0.5) and revealed the weakly first-order nature of the AFM transition, consistent with heat-capacity measurements \cite{ChennpjQuantMat2024,WiedenmannSolidStateCommun1981,BhutaniPRM2020}. Due to the intrinsic $C_3$ symmetry of the honeycomb lattice, there are three possible zigzag domains [ZZ1, ZZ2, and ZZ3 in Fig. \ref{fig1}(d)]. In neutron diffraction, ZZ1, ZZ2, and ZZ3 domains contribute to magnetic Bragg peaks at the three $M$-points of the hexagonal Brillouin zone [$M_1$, $M_2$, and $M_3$ in the inset of Fig. \ref{fig1}(d)]. Therefore, one can use the Bragg peak intensities to directly monitor the populations of these zigzag domains.
	
	\begin{figure*}[t!]
		\centering{\includegraphics[width=0.90\textwidth]{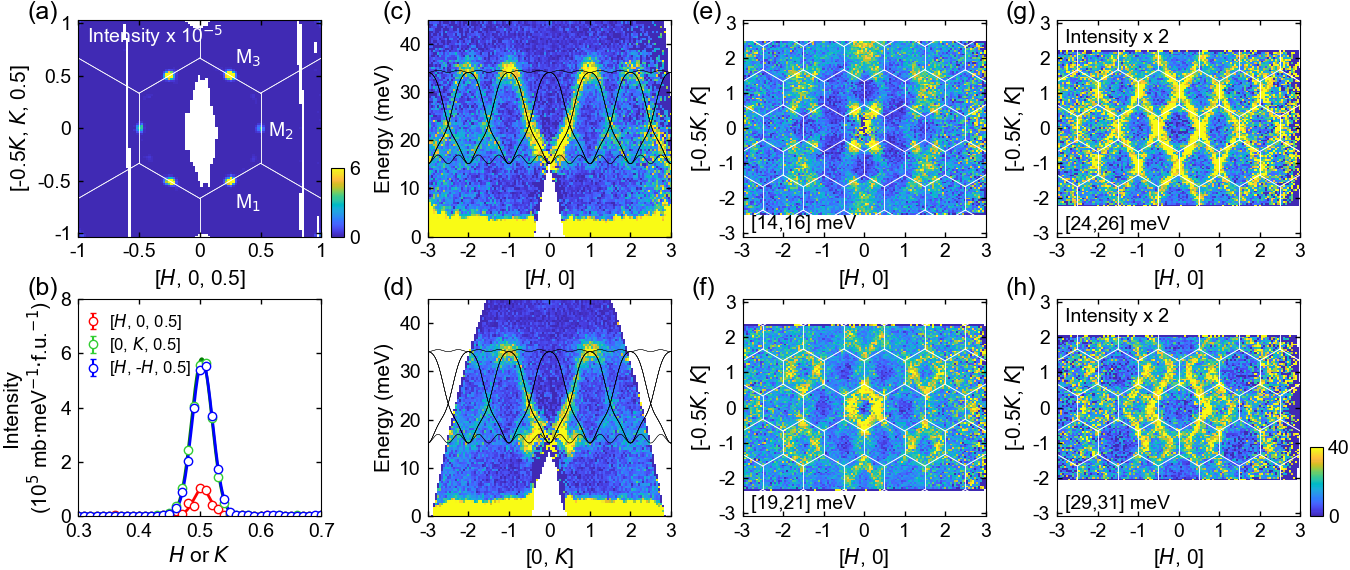}}
		\caption{(a) Elastic scattering pattern of the ($H$, $K$, 0.5) plane at 6.0 K. White solid lines mark Brillouin zone boundaries. f.u.: formula unit. (b) Momentum dependence of the intensities around three magnetic Bragg peaks [M$_2$, M$_3$, and M$_1$ in the inset of Figs. \ref{fig1}(d) and \ref{fig1}(e)] along [$H$, 0, 0.5], [0, $K$, 0.5], and [$H$, -$H$, 0.5]. Solid curves are Gaussian fits. (c) and (d) Magnetic excitation spectra along [$H$, 0] and [0, $K$] at 6.0 K. Black solid curves are calculated spin-wave dispersions using SpinW \cite{TothJPCM2015}. (e)-(h) Constant-energy slices of the spectra around 15 meV, 20 meV, 25 meV, and 30 meV at 6.0 K.}
		\label{fig2}
	\end{figure*}
	
	Recent optical linear dichroism (LD) and elastocaloric measurements have revealed a novel nematic state in FePSe$_3$, in which the $C_3$ lattice symmetry gives rise to a three-state Potts nematicity associated with the underlying AFM order \cite{HwangboNP2024,NiArxiv2023}. Without externally applied uniaxial strain, the three equally populated zigzag domains below $T_{\rm{N}}$ constitute a three-state Potts state. In-situ uniaxial strain was used to tune this three-state Potts state and extract the corresponding susceptibility, which exhibits divergent behavior near $T_{\rm{N}}$, indicating the formation of a three-state Potts nematic state \cite{HwangboNP2024}. Since FePSe$_3$ is an AFM insulator, the nematicity is solely related to the spin degrees of freedom. FePSe$_3$ therefore offers a unique opportunity to study the physics of three-state Potts nematicity arising from spin interactions. While LD and elastocaloric measurements can provide useful information on the three-state Potts nematic phase, they are not microscopic probes and thus cannot unveil the wave vector of the magnetic order or the momentum dependence of the spin dynamics in the nematic state. In copper- and iron-based high-temperature superconductors \cite{HinkovScience2008,LuScience2014}, anisotropic spin excitations in momentum space detected by inelastic neutron scattering (INS) experiments provided direct proof of the nematic state and its connection with anisotropic transport and electronic properties \cite{FernandesAnnuRevCMP2010,NiePNAS2014,DaiRMP2015,BohmerNP2022}.
	
	\begin{figure}[t!]
		\hspace{0cm}\includegraphics[width=7.8cm]{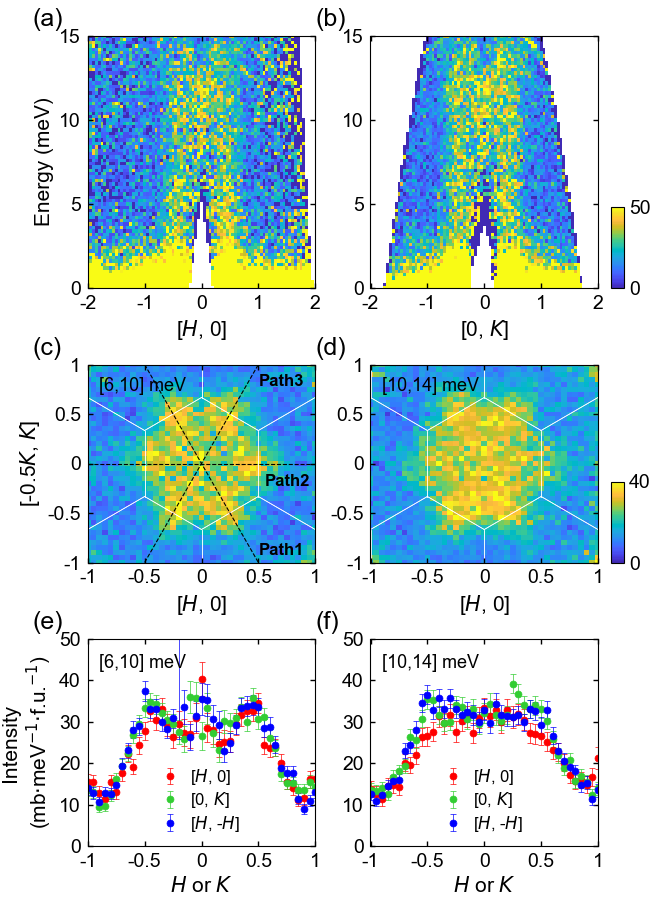}
		\caption{(a) and (b) Spin excitation spectra along [$H$, 0] and [0, $K$] at 108.8 K. (c) and (d) Constant-energy slices of the excitation spectra around 8 meV and 12 meV at 108.8 K. (e) and (f) Momentum dependence of the intensities around 8 meV and 12 meV along [$H$, 0], [0, $K$], and [$H$, -$H$] [paths 2, 3, and 1 in (c), respectively].}
		\label{fig3}
	\end{figure}
	
	\begin{figure}[t!]
		\hspace{0cm}\includegraphics[width=7.8cm]{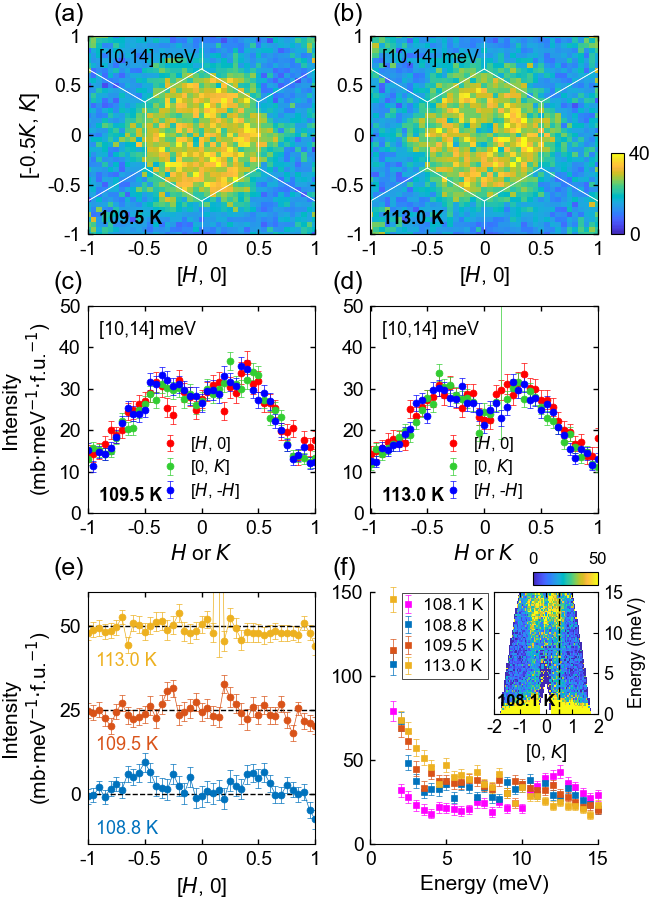}
		\caption{(a) and (b) Constant-energy slices of the spin excitation spectra around 12 meV at 109.5 K and 113.0 K. (c) and (d) Momentum dependence of the intensities around 12 meV along [$H$, 0], [0, $K$], and [$H$, -$H$] at 109.5 K and 113.0 K. (e) Momentum dependence of the intensity difference (see text) around 12 meV. Data from different temperatures are evenly offset for clarity. Horizontal dashed lines indicate zero intensity for each temperature. (f) Energy dependence of the intensities around (0, 0.5) at 108.1 K, 108.8 K, 109.5 K, and 113.0 K. Inset shows the spin excitation spectrum along [0, $K$] at 108.1 K. The vertical dashed line marks (0, 0.5).}
		\label{fig4}
	\end{figure}
	
	To study possible nematic spin excitations in FePSe$_3$, we conducted INS experiments on FePSe$_3$ single crystals under the application of uniaxial strain, which allows direct control over the populations of different zigzag domains. For a magnet with a strongly first-order AFM transition, critical spin fluctuations around $T_{\rm{N}}$ are typically absent, resulting in no or only weak nematic behavior. In contrast, for a weakly first-order AFM transition, uniaxial strain can enhance critical spin fluctuations near the AFM transition regime, thereby promoting a nematic phase, as observed in weakly first-order AFM BaFe$_2$As$_2$ \cite{KimPRB2011,LuPRB2016,LiuNC2020} and strongly first-order AFM SrFe$_2$As$_2$ \cite{TamPRB2019}. In the case of unstrained FePSe$_3$, the weakly first-order nature of the AFM transition \cite{ChennpjQuantMat2024,BhutaniPRM2020} suggests that static magnetic order and spin excitations should obey $C_3$ symmetry, consistent with the underlying honeycomb lattice. However, when uniaxial strain detwins the three magnetic domains, $C_2$-symmetric diffraction and spin-wave patterns are expected below $T_{\rm{N}}$. When the system is warmed above $T_{\rm{N}}$ with suppressed static magnetic order, the spin excitations should become gapless immediately with the $C_3$ symmetry if there are no critical spin fluctuations or a nematic phase. Conversely, if the restoration of the $C_3$ symmetry is delayed until several K above $T_{\rm{N}}$, the $C_2$-symmetric spin excitations in the paramagnetic state would be a signature of the underlying three-state Potts nematicity appearing at temperatures above $T_{\rm{N}}$.
	
	We utilized a novel uniaxial strain device previously used to detwin FeSe single crystals, in which thin aluminum plates are mounted onto the invar alloy frames to induce weakly temperature-dependent uniaxial tensile strain \cite{LiuNC2025}. About 1 gram of FePSe$_3$ single crystals were co-aligned and attached to the thin aluminum plates of the strain device, with the crystallographic ($H$, 0, $L$) plane placed in the horizontal scattering plane [Fig. \ref{fig1}(c)]. When cooling down to below 150 K, a uniaxial tensile strain of about 0.6\% is applied along the $[-0.5K, K, 0]$ direction ($i.e.$, perpendicular to the nearest Fe-Fe bond) [Fig. \ref{fig1}(f)], which was determined by in-situ strain-gauge measurements \cite{SM,MoCPL2024}. Our INS experiment using this strain device was conducted on the time-of-flight spectrometer MERLIN \cite{BewleyPhysicaB2006} at the ISIS neutron and muon source, Rutherford Appleton Laboratory, UK. More details of the experiment and data analysis can be found in \cite{SM}.
	
	Figure \ref{fig1}(g) presents the temperature dependence of the intensity at the magnetic Bragg peak $M_2$ (0, 0.5, 0.5) under uniaxial strain, which shows a typical order-parameter behavior. By fitting the data with a power-law function $I = A (T - T_{\rm{N}})^{2\beta} + B$ ($A$ and $B$ are the scale and background constants, respectively, and $\beta$ is the critical exponent of the order parameter), we find $T_{\rm{N}} = 108.59(2)$ K and $\beta = 0.10(1)$. The obtained $T_{\rm{N}}$ coincides with the sharp magnetic susceptibility anomaly in the unstrained case [Fig. \ref{fig1}(g)], which confirms that the transition temperature is unchanged by the uniaxial strain. However, the critical exponent becomes closer to the limit of the 2D Ising model ($\beta = 1/8$) \cite{ChennpjQuantMat2024}, suggesting the magnetic phase transition approaches second order under uniaxial strain, similar to previous works on BaFe$_2$As$_2$ \cite{KimPRB2011,LuPRB2016,LiuNC2020}.
	
	A diffraction pattern of the ($H$, $K$, 0.5) plane at 6.0 K is shown in Fig. \ref{fig2}(a), from which we find that $M_1$ and $M_3$ basically have the same intensity while $M_2$ is much weaker. Since our sample consists of many co-aligned single crystals [Fig. \ref{fig1}(c)] \cite{SM}, we expect the three zigzag domains to be present in equal quantities without the strain \cite{ChennpjQuantMat2024}. Our observation thus suggests that the ZZ2 domain is strongly suppressed by the uniaxial strain, and the ZZ1 and ZZ3 domains are favored [see Fig. \ref{fig1}(e)]. Such strain control of the zigzag domains is consistent with previous reports based on optical methods \cite{HwangboNP2024,NiArxiv2023}. By extracting the integrated Bragg peak intensities of $M_1$, $M_2$, and $M_3$ ($I_{\rm{M1}}$, $I_{\rm{M2}}$, and $I_{\rm{M3}}$) [Fig. \ref{fig2}(b)], we estimate a detwinning ratio of $\eta = \frac{(I_{\rm{M1}}+I_{\rm{M3}})/2-I_{\rm{M2}}}{(I_{\rm{M1}}+I_{\rm{M3}})/2+I_{\rm{M2}}} \approx 75.7\%$, which is comparable with the experiment on FeSe \cite{LiuNC2025}.
	
	Next, we turn to the strain effect on the magnetic excitation spectra. Since FePSe$_3$ is a quasi-2D antiferromagnet with interlayer exchange interactions much weaker than intralayer ones \cite{ChennpjQuantMat2024}, we integrated over a wide range of $L$ and present the spectra within the honeycomb plane \cite{SM}. Spin waves along the [$H$, 0] and [0, $K$] directions at 6.0 K are shown in Figs. \ref{fig2}(c) and \ref{fig2}(d), respectively. While the overall dispersions are similar for the two directions, the intensity distributions (dynamic structure factors) exhibit a clear difference. In particular, the low-lying spin-wave branch around 16 meV is strongly suppressed along [$H$, 0], suggesting that this branch is associated with ZZ2 domain \cite{SM} and that low-energy spin excitations are more susceptible to external perturbations. Nevertheless, the measured dispersions can still be well described by previously determined parameters within linear spin-wave theory [Figs. \ref{fig2}(c) and \ref{fig2}(d)] \cite{ChennpjQuantMat2024}, indicating that the exchange interactions remain essentially unchanged under $\sim$0.6\% tensile strain and that zigzag domain repopulation is the dominant effect. Constant-energy slices in Figs. \ref{fig2}(e)-(h) further demonstrate that the system exhibits $C_2$ symmetry at low temperature, consistent with the pronounced optical LD signals \cite{HwangboNP2024,NiArxiv2023}.
	
	On warming to 108.8 K, which is slightly above $ T_{\rm{N}}$, sharp spin waves completely disappear and the broad excitations shift to lower energies [Figs. \ref{fig3}(a) and \ref{fig3}(b)]. Despite very similar spectra along the [$H$, 0] and [0, $K$] directions, spin excitations along the [$H$, 0] direction are overall broader than those along the [0, $K$] direction. Figures \ref{fig3}(c) and \ref{fig3}(d) present the constant-energy slices taken around 8 meV and 12 meV, respectively. While the intensity distribution in the hexagonal Brillouin zone shows nearly $C_3$ symmetry around 8 meV, violation of it is evident for the pattern around 12 meV, which is $C_2$-symmetric about the $c$ axis. The breaking of the $C_3$ symmetry can be more clearly seen from the momentum dependence of the intensity, as shown in Figs. \ref{fig3}(e) and \ref{fig3}(f). Around 12 meV, the intensities along the [0, $K$] and [$H$, -$H$] directions are essentially the same, but they deviate from those along the [$H$, 0] direction near the $M$ points of the Brillouin zone [$H$ or $K = 0.5$ in Fig. \ref{fig3}(f)] by approximately 10 mb$\cdot$meV$^{-1}$$\cdot$f.u.$^{-1}$. These observations suggest the spin fluctuations of FePSe$_3$ show nematicity without the presence of long-range magnetic order.
	
	The $C_3$ symmetry of the spin excitation spectra is gradually restored at higher temperatures (109.5 K and 113.0 K), as can be seen in Figures \ref{fig4}(a)-(d). To show the evolution of the $C_3$ symmetry breaking (or nematicity) above $T_{\rm{N}}$, we use the intensity difference $\Delta I(\textbf{Q})$, which is defined as $\Delta I(\textbf{Q}) = [I_1(\textbf{Q}) + I_3(\textbf{Q})]/2 - I_2(\textbf{Q})$, where $I_1(\textbf{Q})$, $I_2(\textbf{Q})$, and $I_3(\textbf{Q})$ are the intensity distributions along the paths 1, 2, and 3 shown in Fig. \ref{fig3}(c). We note that, for strictly $C_3$-symmetric spin excitations, $\Delta I(\textbf{Q})$ = 0. Therefore, this quantity can measure the $C_3$ symmetry breaking and may be regarded as an effective nematic order parameter \cite{FradkinAnnuRevCMP2010}. The $\Delta I(\textbf{Q})$ at three temperatures is presented in Fig. \ref{fig4}(e), where it is nonzero at 108.8 K and 109.5 K, and finally dies away at 113.0 K. The complete disappearance of the $C_3$ symmetry breaking at 113.0 K also suggests that the feature observed at lower temperatures is not simply due to the applied strain ($ e.g.$, mechanical effects) but is intrinsic to the magnetic response. The quick decay of the nematic signal above $T_{\rm{N}}$ is consistent with the hyperbola behavior (with respect to temperature) of the nematic susceptibility determined previously \cite{HwangboNP2024}. With increasing strain, the AFM transition becomes more second-order-like, thus increasing the critical regime as shown in \cite{HwangboNP2024}. For the level of strain we applied ($\sim$0.6 \%), the nematic and AFM transition temperatures are expected to nearly merge together \cite{HwangboNP2024}, thus giving rise to the narrow temperature regime above $T_{\rm{N}}$ where the $C_3$-symmetry-breaking spin excitations are observed.
	
	To further check the characteristics of the spin excitations near $T_{\rm{N}}$, we show the energy dependence of the intensity at (0, 0.5) in Fig. \ref{fig4}(f). At 108.1 K (slightly below $T_{\rm{N}}$), an energy gap of $\sim$13 meV can be resolved, signifying the Ising anisotropy of the spin waves [see also in the inset of Fig. \ref{fig4}(f)]. However, the energy gap fully closes at 108.8 K and above. This suggests the nematic spin excitations we observed are in the paramagnetic state, which differs from the anisotropic spin waves below $T_{\rm{N}}$ due to the detwinning of zigzag domains. Therefore, the nematic spin excitations slightly above $T_{\rm{N}}$, a hallmark of the three-state Potts nematic state, are a vestigial order for the zigzag AFM order below $T_{\rm{N}}$, and likely arise from strain-induced nematic phase similar to iron pnictides \cite{FernandesAnnuRevCMP2010,NiePNAS2014,DaiRMP2015,BohmerNP2022}. 
	
	Our experimental results unveil the following sequences of the magnetic phases in FePSe$_3$. On cooling from the high-temperature paramagnetic spin-disordered state, the system first enters the uniaxial strain-induced three-state Potts nematic state, where the spin excitations are still highly fluctuating but show anisotropy in reciprocal space due to spin-lattice coupling. In the absence of external strain, the nematic order exhibits equal probability along the three equivalent directions of the honeycomb lattice, which hence leads to the $C_3$-symmetric excitation patterns as observed previously \cite{ChennpjQuantMat2024}. When the external strain is applied to induce a tiny lattice distortion, it disfavors the nematic director perpendicular to the strain direction \cite{HwangboNP2024,NiArxiv2023}. Due to the divergent nature of the nematic susceptibility, the strain effect is most pronounced close to $T_{\rm{N}}$ \cite{HwangboNP2024}, which results in the $C_2$-symmetric excitation pattern. At temperatures lower than $T_{\rm{N}}$, the long-range zigzag AFM order condenses mainly around the $M$-points where the nematic fluctuations are most significant, and finally establishes the two preferred zigzag domains of dominant population.
	
	The nematic spin excitations observed in FePSe$_3$ are reminiscent of those in iron-based superconductors such as BaFe$_{\rm{2-x}}$Ni$_{\rm{x}}$As$_2$ \cite{LuScience2014} and FeSe \cite{LuNP2022,LiuNC2025}. In these systems, the $C_4$ symmetry of the paramagnetic spin excitations becomes $C_2$ symmetry in the magnetic ordered state due to orthorhombic lattice distortion and collinear AFM order along the $a$ axis of the orthorhombic lattice \cite{DaiRMP2015}. On warming to the paramagnetic tetragonal state slightly above $T_{\rm{N}}$, spin excitations are $C_4$-symmetric without uniaxial strain. However, when a uniaxial strain is applied along one of the tetragonal axis, the low-energy spin excitations become anisotropic along the [$H$, 0] and [0, $K$] directions in the 2D square Brillouin zone over a broad range above $T_{\rm{N}}$, forming a strain-driven nematic phase \cite{LuNP2022,LiuNC2025,LuScience2014}. In contrast, the uniaxial strain induced spin excitation anisotropy in FePSe$_3$ form in an extremely narrow regime above $T_{\rm{N}}$ before becoming isotropic in reciprocal space. This is most likely due to the fact that the AFM phase transition under uniaxial strain is still first-order-like with a narrow critical temperature regime and weak critical spin fluctuations. In addition, while the nematic phase in iron-based superconductors is clearly associated with tetragonal-to-orthorhombic lattice distortion \cite{FernandesAnnuRevCMP2010,NiePNAS2014,DaiRMP2015,BohmerNP2022}, there is no known lattice distortion associated with the AFM phase transition in FePSe$_3$ \cite{WiedenmannSolidStateCommun1981,ChennpjQuantMat2024}, suggesting that magnetoelasticity may not drive the order-disorder phase transition \cite{LanePRB2025}. Nevertheless, the fact that a uniaxial tensile strain of about 0.6\% induces anisotropy in paramagnetic spin excitations of FePSe$_3$ suggests a small but finite magnetoelastic coupling.
	
	In summary, by performing INS experiments on FePSe$_3$ single crystals under uniaxial tensile strain, we have mapped out the spin excitation spectra and studied their evolution with temperature. Although the applied strain is only about $\sim$0.6\%, we observed a significant detwinning effect of the zigzag domains in the AFM ordered state and the corresponding spin waves featuring $C_2$ symmetry. Moreover, in the paramagnetic state just above the ordering temperature, the spin excitations still exhibit broken $C_3$ symmetry in a narrow temperature regime, suggesting the presence of magnetoelastic coupling and the underlying three-state Potts nematicity in the spin excitation spectra of FePSe$_3$.
	\newline 
	
	The neutron scattering work at Rice is supported by the United States Department of Energy, Basic Energy Sciences, DE-SC0012311 (P.D.). The single-crystal growth work at Rice is supported by the Robert A. Welch Foundation under grant no. C-1839 (P.D.). X.L. at BNU is supported by the Scientific Research Innovation Capability Support Project for Young Faculty (Grant No. 2251300009). The experiment at the ISIS Neutron and Muon Source was supported by a beam time allocation RB2510117 from the Science and Technology Facilities Council.
	
	\bibliographystyle{apsrev4-1}
	\bibliography{FePSe3_strain_reference}

\begin{thebibliography}{48}
\bibitem [{not()}]{note}%
  \BibitemOpen
  \href@noop {} {}\bibinfo {note} {In two-dimensional systems of interest, the
  $C_6$ group can be generated by the $C_3$ and $C_2$ subgroups. Since the
  nematic state retains $C_2$ symmetry, the distinctions between $C_3$ and
  $C_6$ symmetries become irrelevant, rendering them effectively equivalent in
  this context.}\BibitemShut {Stop}%
\end{thebibliography}
	
	\pagebreak
	\pagebreak

	\widetext
	
	\begin{center}
		\textbf{\large Supplemental Material for ``Signatures of three-state Potts nematicity in spin excitations of the van der Waals antiferromagnet FePSe$_3$''}
	\end{center}
	
	\setcounter{equation}{0}
	\setcounter{table}{0}
	\setcounter{page}{1}
	\makeatletter
	\renewcommand{\theequation}{S\arabic{equation}}
	\renewcommand{\thefigure}{S\arabic{figure}}
	
	\section{FePSe$_3$ single crystal growth and characterization}
\begin{figure}[b!]
		\hspace{0cm}\includegraphics[width=16cm]{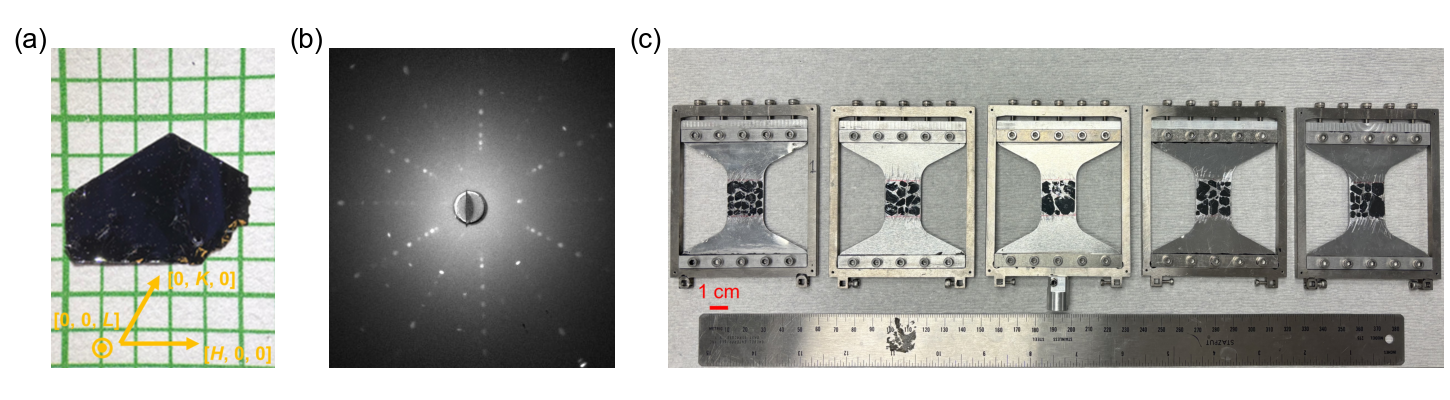}
		\caption{(a) One piece of representative FePSe$_3$ single crystal against a millimeter grid. (b) X-ray Laue diffraction pattern taken on the (0, 0, $L$) surface. Photograph of the disassembled strain device with the FePSe$_3$ single crystals attached.}
		\label{figs1}
	\end{figure}

	High-quality FePSe$_3$ single crystals were synthesized using a chemical vapor transport method. A mixture of 0.52 g iron powder, 0.29 g phosphorus chunks, 2.20 g selenium powder, and 0.25 g iodine was loaded into a quartz tube (outer diameter 35 mm, wall thickness 1.5 mm), which was then evacuated and sealed under vacuum. The sealed portion of the tube was approximately 17 cm in length. The tube was placed in a one-zone furnace with the starting materials positioned at the hot end. Since the central region of the furnace has nearly uniform temperature, the tube was shifted 8 cm away from the center to create a temperature gradient. The furnace was programmed to ramp to 620 $^{\circ}$C within 6 hours, held at this temperature for 108 hours, and then switched off. Large FePSe$_3$ single crystals with a hexagonal morphology were obtained at the cold end of the tube [Fig. \ref{figs1}(a)]. X-ray Laue diffraction [Fig. \ref{figs1}(b)] confirmed that the edge of the hexagonal crystal is perpendicular to the [$H$, 0, 0] direction. For inelastic neutron scattering (INS) experiments, the FePSe$_3$ crystals were glued to thin aluminum plates using hydrogen-free CYTOP adhesive. These plates were mounted on invar alloy frames [Fig. \ref{figs1}(c)], where the mismatch in thermal expansion coefficients between the invar alloy frame and the aluminum plate applied a uniaxial tensile strain to the samples upon cooling \cite{LiuNC2025}.
	
	The magnitude of the uniaxial tensile strain applied to the FePSe$_3$ samples was determined using a custom-built cryogenic digital image correlation (CDIC) system \cite{MoCPL2024,LiuNC2025}. A piece of FePSe$_3$ single crystal was mounted on a compact strain device of the same design, which is compatible with a low-temperature cryostat equipped with an optical window. The relative displacements between feature points on the sample were tracked during cooling. The effective strain [Fig. 1(f)] was calculated from the difference in distance shifts along and perpendicular to the aluminum plate. More details of this measurement can be found in \cite{LiuNC2025} and \cite{MoCPL2024}.
	
\begin{figure}[t!]
		\hspace{0cm}\includegraphics[width=16cm]{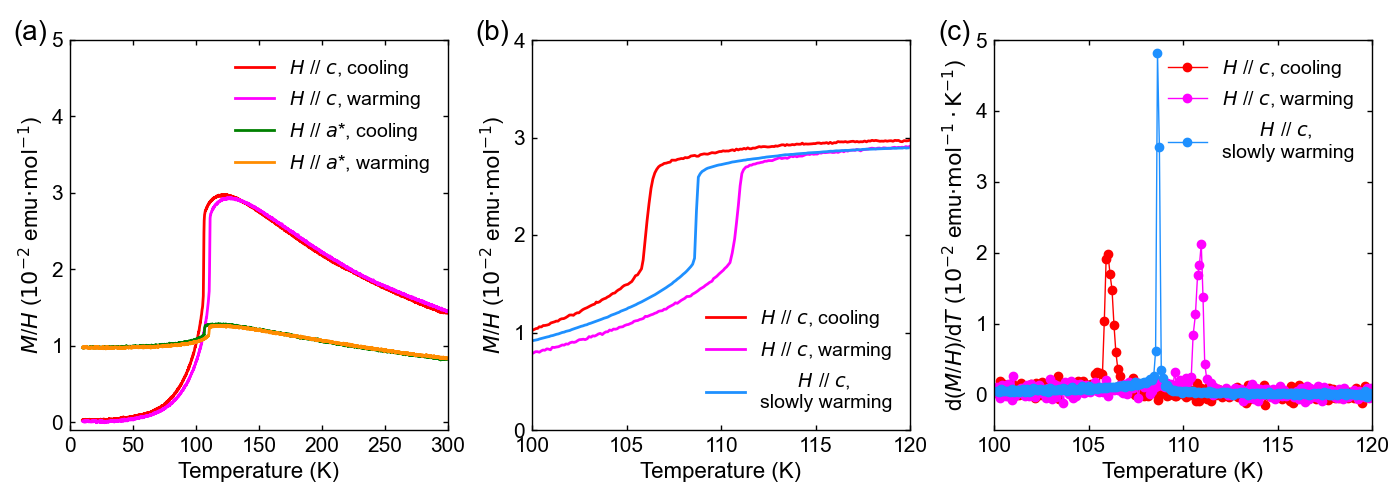}
		\caption{(a) Temperature dependence of the magnetic susceptibility under a 0.5 T field applied along the $c$ axis and the $\mathbf{a}^{\ast}$ direction, showing both warming and cooling processes. (b) Enlarged view of the magnetic susceptibility between 100 K and 120 K for the field applied along the $c$ axis. The magnetic susceptibility measured with a much slower temperature sweep rate is also shown. (c) The temperature derivative of the magnetic susceptibility shown in (b).}
		\label{figs2}
	\end{figure}

	To examine the antiferromagnetic (AFM) phase transition in our FePSe$_3$ single crystals, we measured the magnetic susceptibility using a Quantum Design PPMS. Fig. \ref{figs2}(a) displays the temperature dependence of the susceptibility under a 0.5 T field applied along both the $c$ axis and the $\mathbf{a}^{\ast}$ direction, with a temperature sweep rate of 3 K/min for both cooling and warming. The results are generally consistent with previous reports \cite{NiArxiv2023,BasnetAEM2024} and indicate an easy-axis spin configuration in the AFM ordered state. A clear thermal hysteresis is observed, as the cooling and warming curves do not overlap. This feature is highlighted in the enlarged view in Fig. \ref{figs2}(b). The N\'eel temperatures ($T_{\rm{N}}$), defined by the peak positions of the temperature derivative of the susceptibility [Fig. \ref{figs2}(c)], are $\sim$106.04 K on cooling and $\sim$110.97 K on warming. However, when the sweep rate was reduced to 0.05 K/min, the peak position of d($M$/$H$)/d$T$ shifted to $\sim$108.65 K, essentially midway between the two faster-sweep values. This result demonstrates that an accurate determination of $T_{\rm{N}}$ requires a slow sweep rate to ensure thermal equilibrium. Notably, the $T_{\rm{N}}$ obtained from our neutron diffraction measurement [Fig. 1(g)], 108.59(2) K, agrees very well with the thermally equilibrated result. That slow-sweep measurement took approximately 4 hours to cover the temperature range from $\sim$100 K to $\sim$117 K.

	\section{Inelastic neutron scattering experiment and additional data}
	
	Our INS experiment was performed with the MERLIN spectrometer. To avoid the large background from the invar frames of the strain device, these frames were covered with cadmium foils before loading into the spectrometer. With MERLIN's repetition-rate multiplication mode, data from three incident neutron energies ($E_i$ = 61.9 meV, 21.3 meV, and 11.1 meV) were simultaneously collected. The chopper frequency was set to be 250 Hz. The energy resolutions at zero energy transfer are about 3.5 meV, 0.85 meV, and 0.35 meV for $E_i$ = 61.9 meV, 21.3 meV, and 11.1 meV, respectively. We measured at six temperatures (6.0 K, 100.0 K, 108.1 K, 108.8 K, 109.6 K, and 113.0 K) with the sample rotated for 60 degrees along the [-0.5$K$, $K$, 0] direction.
	
	\begin{figure}[h!]
		\hspace{0cm}\includegraphics[width=10cm]{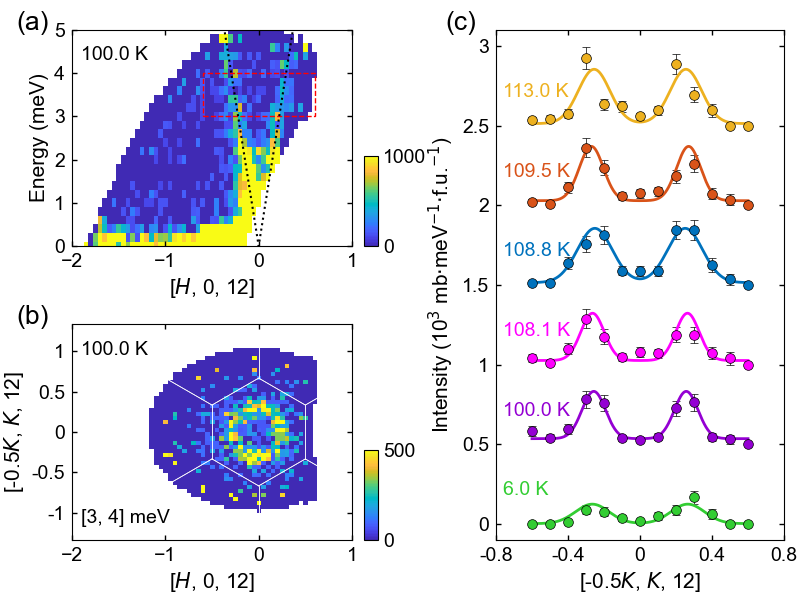}
		\caption{(a) Acoustic phonon emerging from (0, 0, 12) at 100.0 K. Black dashed lines indicate the linear phonon dispersion. (b) Constant-energy slice of the ($H$, $K$, 12) plane around 3.5 meV showing the acoustic phonon. (c) Momentum dependence of the phonon intensity around 3.5 meV along the [-0.5$K$, $K$, 12] direction [red dashed box in (a)] for six temperatures. Data are vertically offset by 500 mb$\cdot$meV$^{-1}\cdot$f.u.$^{-1}$ for clarity. Solid curves are double-Gaussian fits to the data.}
		\label{figs3}
	\end{figure}
	
	Data reduction was carried out using Mantid \cite{Arnold2014} and HORACE \cite{Ewings2016}. Except for the diffraction data in Fig. 1(g), all presented intensities were converted to absolute units of mb$\cdot$meV$^{-1}\cdot$f.u.$^{-1}$ using the phonon scattering around (0, 0, 12) (see Fig. \ref{figs3}) following a standard procedure \cite{XuRSI2013}. The momentum- and/or energy-integration ranges and the incident energies ($E_i$s) for the presented data are summarized in Table \ref{tb1}.
	
\begin{figure}[b!]
		\hspace{0cm}\includegraphics[width=16.5cm]{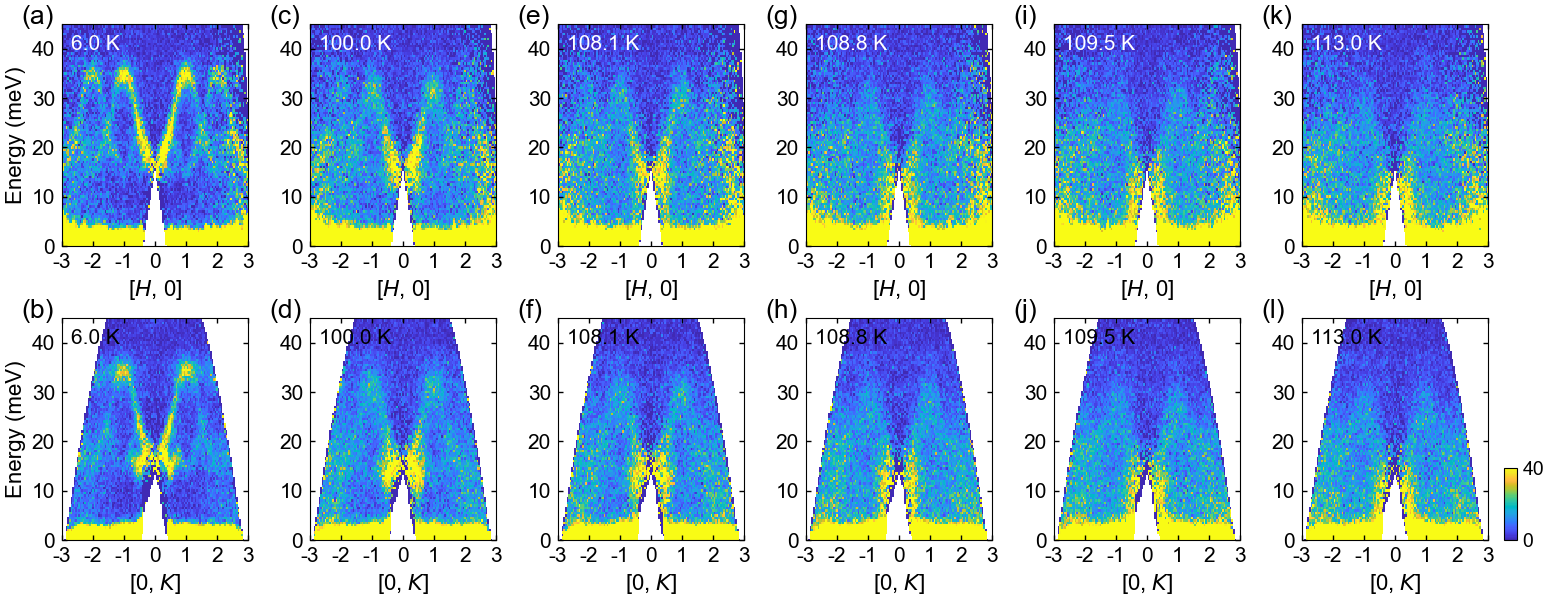}
		\caption{High-energy spin excitation spectra at six temperatures along [$H$, 0] [(a), (c), (e), (g), (i), and (k)] and [0, $K$] [(b), (d), (f), (h), (j), and (l)].}
		\label{figs4}
	\end{figure}

\begin{figure}[t!]
		\hspace{0cm}\includegraphics[width=14cm]{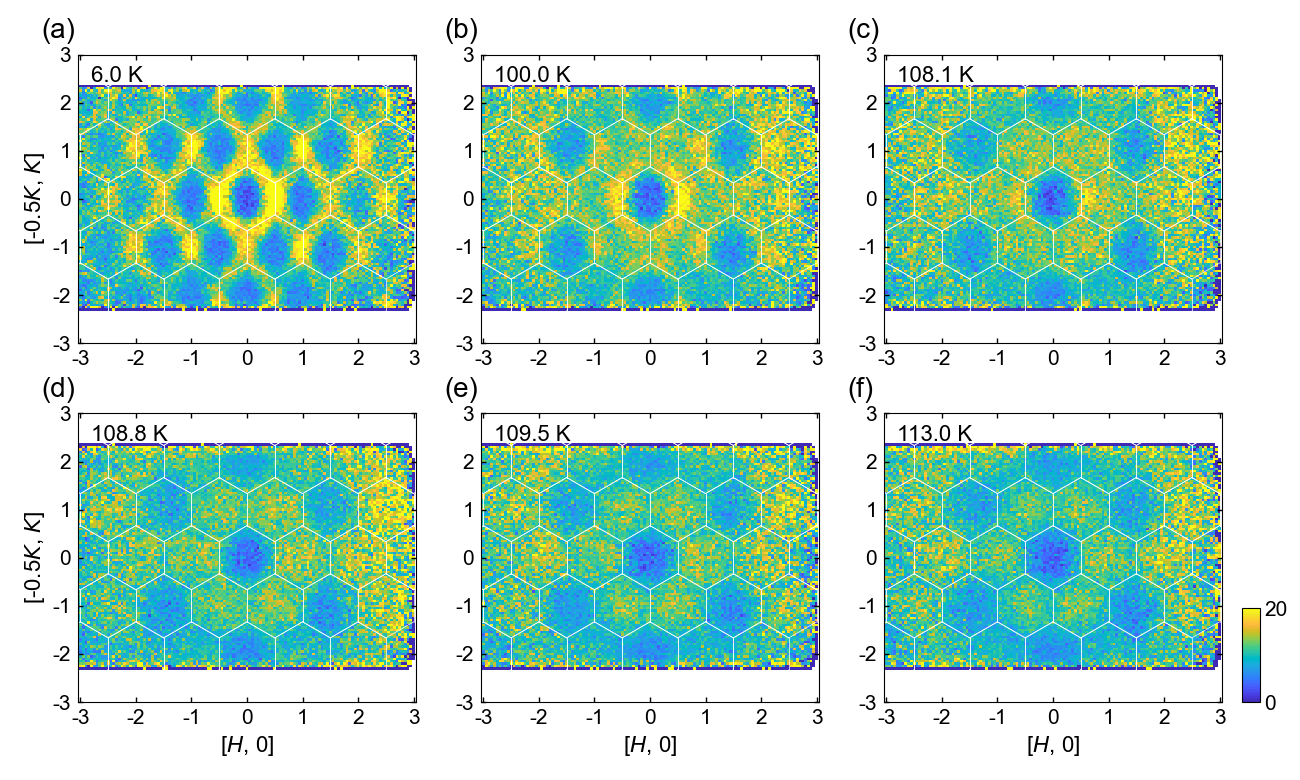}
		\caption{Constant-energy slices of the spin excitation spectra around 25 meV at six temperatures.}
		\label{figs5}
	\end{figure}

\begin{figure}[t!]
		\hspace{0cm}\includegraphics[width=16.5cm]{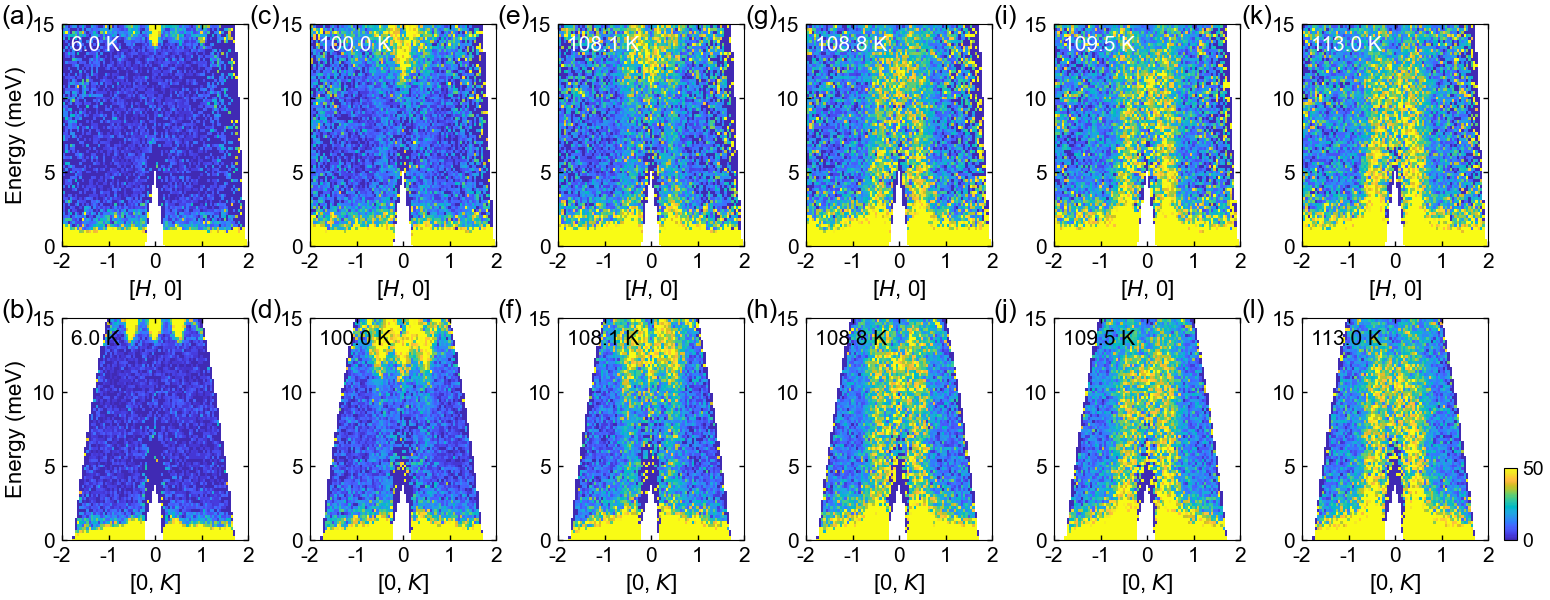}
		\caption{Low-energy spin excitation spectra at six temperatures along [$H$, 0] [(a), (c), (e), (g), (i), and (k)] and [0, $K$] [(b), (d), (f), (h), (j), and (l)].}
		\label{figs6}
	\end{figure}

	Fig. \ref{figs4} shows the high-energy spin excitation spectra at six temperatures. Dispersive spin waves are sustained up to 108.1 K, whereas no clear dispersion is discernible at 108.8 K and above. The symmetry of the excitation patterns is examined in Fig. \ref{figs5}, which displays constant-energy slices around 25 meV. The thermal evolution of the low-energy excitations is shown in Fig. \ref{figs6}.

	\section{Spin waves from different zigzag antiferromagnetic domains}
\begin{figure}[t!]
		\hspace{0cm}\includegraphics[width=12cm]{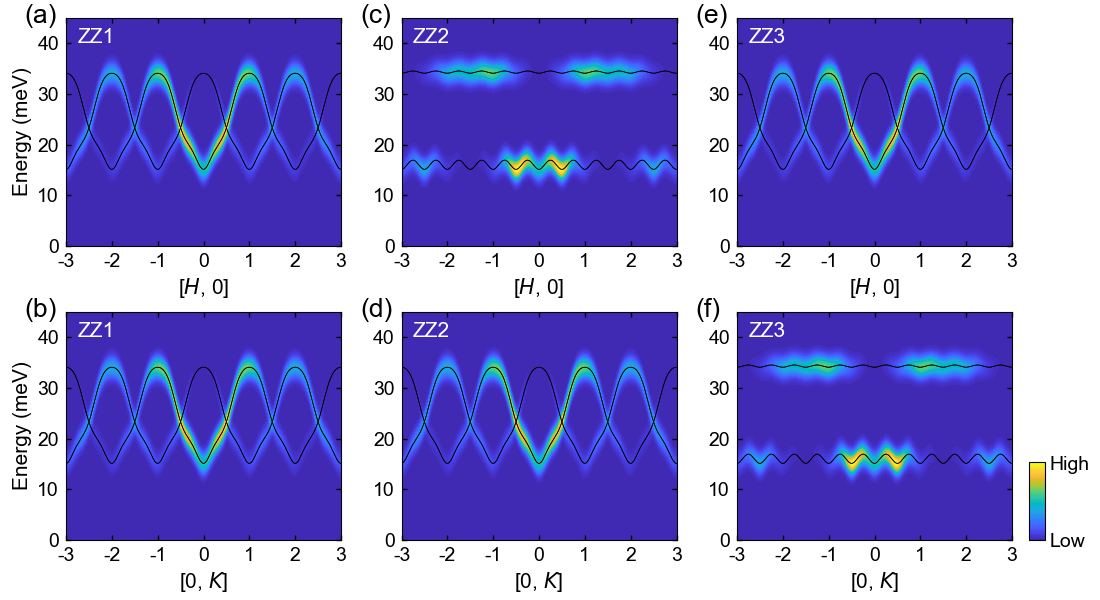}
		\caption{Calculated spin-wave spectra for the three zigzag domains (ZZ1, ZZ2, and ZZ3) along [$H$, 0] and [0, $K$]. The black solid curves are the corresponding spin-wave dispersions.}
		\label{figs7}
	\end{figure}

\begin{figure}[h!]
		\hspace{0cm}\includegraphics[width=9cm]{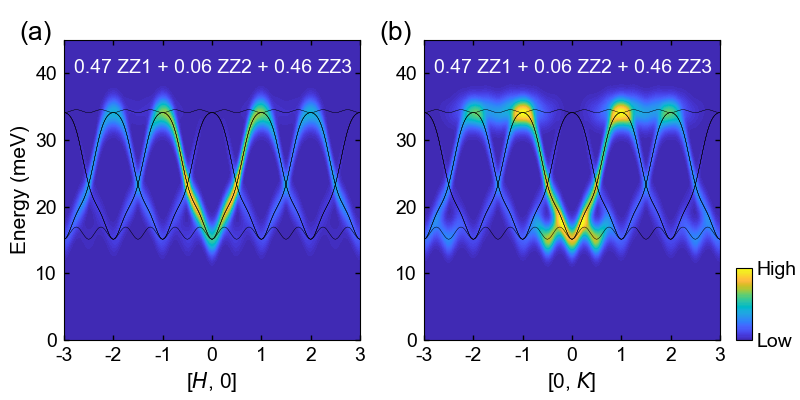}
		\caption{Calculated spin-wave spectra along [$H$, 0] and [0, $K$] based on the domain population determined from the diffraction measurement. The black solid curves are the corresponding spin-wave dispersions.}
		\label{figs8}
	\end{figure}

	For unstrained FePSe$_3$, three zigzag domains coexist in the AFM ordered state [ZZ1, ZZ2, and ZZ3 as shown in Fig. 1(d)]. The spin waves along a given trajectory generally receive contributions from all three domains. Figure \ref{figs7} shows the calculated spin-wave spectra for ZZ1, ZZ2, and ZZ3 individually. The observed spectrum is a linear combination of these, weighted by the domain population. In our experiment, uniaxial strain strongly suppresses ZZ2 while favoring ZZ1 and ZZ3. Diffraction measurements [Fig. 2(a) and (b)] yield a domain ratio of ZZ1 : ZZ2 : ZZ3 $\approx$ 0.47 : 0.06 : 0.46. Using this distribution, we calculated the spin-wave spectra along the [$H$, 0] and [0, $K$] directions (Fig. \ref{figs8}), which reproduce the measured spectra in Fig. 2(c) and (d) with good agreement.
	
	\clearpage
	\begin{table}[t]
		\caption{Momentum- and/or energy-integration ranges and $E_i$s for the data presented in figures.}
		\begin{ruledtabular}
			\begin{tabular}{cccc}
				Data &\makecell{Momentum-integration\\range (r.l.u.)}& \makecell{Energy-integration\\range (meV)}& $E_i$ (meV) \\
				\midrule
				Fig. 2(a) &\makecell{[-0.05, 0.05] along [0, 0, $L$]}&[-0.5, 0.5]&11.1\\
				\hline
				Fig. 2(b) &\makecell{[-0.1, 0.1] along [0.5$K$, -$K$, 0] (red)\\ $$[-0.1, 0.1] along [-$H$, 0.5$H$, 0] (green) \\ $$[-0.1, 0.1] along [0.5$K$, 0.5$K$, 0] (blue)\\ $$ [-0.05, 0.05] along [0, 0, $L$]}&[-0.5, 0.5]&11.1\\
				\hline
				Fig. 2(c)&\makecell{[-0.1, 0.1] along [0.5$K$, -$K$, 0] \\ $$[-10, 10] along [0, 0, $L$]}&-&61.9\\
				\hline
				Fig. 2(d)&\makecell{[-0.1, 0.1] along [-$H$, 0.5$H$, 0] \\ $$[-10, 10] along [0, 0, $L$]}&-&61.9\\
				\hline
				Fig. 2(e)-(h)&\makecell{[-10, 10] along [0, 0, $L$]}&[-1, 1]&61.9\\
				\hline
				Fig. 3(a)&\makecell{[-0.1, 0.1] along [0.5$K$, -$K$, 0] \\ $$[-6, 6] along [0, 0, $L$]}&-&21.3\\
				\hline
				Fig. 3(b)&\makecell{[-0.1, 0.1] along [-$H$, 0.5$H$, 0] \\ $$[-6, 6] along [0, 0, $L$]}&-&21.3\\
				\hline
				Fig. 3(c) and (d)&\makecell{[-6, 6] along $[0, 0, L]$}&[-2, 2]&21.3\\
				\hline
				Fig. 3(e) and (f)&\makecell{[-0.1, 0.1] along [0.5$K$, -$K$, 0] (red)\\ $$[-0.1, 0.1] along [-$H$, 0.5$H$, 0] (green) \\ $$[-0.1, 0.1] along [0.5$K$, 0.5$K$, 0] (blue)\\ $$ [-6, 6] along [0, 0, $L$]}&[-2, 2]&21.3\\
				\hline
				Fig. 4(a) and (b)&\makecell{[-6, 6] along $[0, 0, L]$}&[-2, 2]&21.3\\
				\hline
				Fig. 4(c) and (d)&\makecell{[-0.1, 0.1] along [0.5$K$, -$K$, 0] (red)\\ $$[-0.1, 0.1] along [-$H$, 0.5$H$, 0] (green) \\ $$[-0.1, 0.1] along [0.5$K$, 0.5$K$, 0] (blue)\\ $$ [-6, 6] along [0, 0, $L$]}&[-2, 2]&21.3\\
				\hline
				Fig. 4(f)&\makecell{[-0.1, 0.1] along [$H$, 0, 0] \\$$[-0.1, 0.1] along [0.5$K$, -$K$, 0] \\ $$[-6, 6] along [0, 0, $L$]}&-&21.3\\
				\hline
				Fig. 4(f) inset&\makecell{[-0.1, 0.1] along [-$H$, 0.5$H$, 0] \\ $$[-6, 6] along [0, 0, $L$]}&-&21.3\\
				\hline
				Fig. \ref{figs3}(a)&\makecell{[-0.1, 0.1] along [0.5$K$, -$K$, 0] \\ $$[-0.25, 0.25] along [0, 0, $L$]}&-&11.1\\
				\hline
				Fig. \ref{figs3}(b)&\makecell{[-0.25, 0.25] along $[0, 0, L]$}&[-0.5, 0.5]&11.1\\
				\hline
				Fig. \ref{figs3}(c)&\makecell{[-0.1, 0.1] along [0.5$K$, -$K$, 0] \\ $$[-0.25, 0.25] along [0, 0, $L$]}&[3, 4]&11.1\\
				\hline
				\makecell{Fig. \ref{figs4}(a), (c), (e),\\ (g), (i), and (k)}&\makecell{[-0.1, 0.1] along [0.5$K$, -$K$, 0] \\ $$[-10, 10] along [0, 0, $L$]}&-&61.9\\
				\hline
				\makecell{Fig. \ref{figs4}(b), (d), (f),\\ (h), (j), and (l)}&\makecell{[-0.1, 0.1] along [-$H$, 0.5$H$, 0] \\ $$[-10, 10] along [0, 0, $L$]}&-&61.9\\
				\hline
				Fig. \ref{figs5}&\makecell{[-10, 10] along $[0, 0, L]$}&[-5, 5]&21.3\\
				\hline
				\makecell{Fig. \ref{figs6}(a), (c), (e),\\ (g), (i), and (k)}&\makecell{[-0.1, 0.1] along [0.5$K$, -$K$, 0] \\ $$[-6, 6] along [0, 0, $L$]}&-&21.3\\
				\hline
				\makecell{Fig. \ref{figs6}(b), (d), (f),\\ (h), (j), and (l)}&\makecell{[-0.1, 0.1] along [-$H$, 0.5$H$, 0] \\ $$[-6, 6] along [0, 0, $L$]}&-&21.3\\
				
			\end{tabular}	
		\end{ruledtabular}
		\label{tb1}
	\end{table}
	
	\clearpage

\end{document}